\documentclass[12pt,a4paper]{article}
\usepackage{amsmath}
\usepackage{amsfonts}
\usepackage{amssymb}
\usepackage{graphicx}
\usepackage{hyperref}
\usepackage{color,soul}
\usepackage{authblk}

\title{Coherent perfect absorption mediated enhancement of transverse spin in a gap plasmon guide}

\author[1]{Samyobrata Mukherjee}
\author[1,*]{S. Dutta Gupta}

\affil[1]{School of Physics, University of Hyderabad, Hyderabad 500046, Telangana, India}

\affil[*]{Corresponding author: sdghyderabad@gmail.com}

\begin{document}
	\maketitle

\begin{abstract}
We consider a symmetric gap plasmon guide (a folded Kretschmann configuration) supporting both symmeric and antisymmetric coupled surface plasmons. We calculate the transverse spin under illumination from both the sides like in coherent perfect absorption (CPA), whereby all the incident light can be absorbed to excite one of the modes of the structure. Significant enhancement in the transverse spin is shown to be possible when the CPA dip and the mode excitation are at the same frequency. The enhancement results from CPA-mediated total transfer of the incident light to either of the coupled modes and the associated large local fields. The effect is shown to be robust against small deviations from the symmetric structure. The transverse spin is localized in the structure since in the ambient dielectric there are only incident plane waves lacking any structure. 
\end{abstract}

\section{Introduction}

Spin-orbit interaction using light fields has been one of the most researched areas in optics for past two decades \cite{allen1992,allen2000,aiello2009,bliokhreview1,bliokhreview2,aielloreview}. There are still fundamental issues associated with the true nature of spin in optical fields \cite{belinfante1940}. In the recent past there has been a great deal of interest in Bellinfante's elusive transverse spin  in optical systems. Various schemes, mostly theoretical in nature, have been proposed in order to have a deeper understanding of the physical origin and the manifestations (in terms of forces exerted on a particle) and their experimental observation \cite{bliokh2012, bliokh2014,bliokh2015,bekshaev2015,aiello2015,alizadeh2016,jacob2016,zayats2014}. 
It is now well understood that the extraordinary spin can be helicity-dependent as in the case of evanescent waves in total internal reflection of circularly polarized light \cite{bliokh2014}. There is the other variant which is helicity-independent, occurring in systems supporting surface plasmons at metal-dielectric interfaces excited by TM-polarized light \cite{bliokh2012}. In such structures the origin of the transverse spin can be traced to the phase lag of the orthogonal components of the electric field  leading to the ``polarization ellipse" and rotation of the field vector in the plane of incidence. The structure of the spin angular momentum (SAM) density has been studied by Berry \cite{berry2009,berry2015} in dealing with the optical current embodied by the Poynting vector. It was shown how the current can be broken up into the orbital (canonical) and the spin part. The SAM density was shown to be proportional to $Im(\mathbf{E}^*\times\mathbf{E})$ from a direct analogy with the quantum treatment of spin-one particles. A recent breakthrough was the experimental observation of the transverse spin for both helicity-independent \cite{prl_expt} and helicity-dependent \cite{nat_expt} cases. While the former was an indirect measurement with focused structured beam, the latter was a direct one using a nano-cantilever reporting a pressure force of the order of a femtonewton. The nature of the force being tiny explains why the Bellinfante spin is referred to as an elusive entity. The other key feature that emerged from the above studies is the realization that structured light such as interfering plane waves, evanescent waves and surface plasmons are essential for generating transverse spin \cite{bliokh2012,bliokh2014,bliokh2015, bekshaev2015,aiello2015,alizadeh2015}. It is thus a pertinent issue to find novel systems which offer structured light leading to extraordinary spin and to probe ways and means to enhance the elusive spin. It has recently been shown that localized plasmons and Mie resonances can be used as a tool to enhance transverse spin \cite{nirmalya2016}.
\par
In this paper we describe a system that offers structured light fields carrying this extraordinary spin and also demonstrate a means of enhancing the spin. We study a symmetric gap plasmon guide with two thin metal films separated by a dielectric gap. Similar gap plasmon guides have been explored, for example, for fast and slow light \cite{sdg2009}, for selective mode excitation \cite{sdgthesis} and also in the context of quatum plasmonics \cite{sdg2014} (an extension to higher order interference with single photon sources led to the possibility of Hong-Ou-Mandel dip \cite{hong1987} with 100\% visibility). For sufficiently small thickness of the gap the surface plasmons on the two inner dielctric-metal interfaces can get coupled leading to the symmetric and antisymmetric modes \cite{raetherbook}. We focus on the fundamental TM mode which bifurcates into the two branches (symmetric and antisymmetric) under strong coupling (smaller gap). The field inside the gap is highly structured with the necessary phase lag between the two non-vanishing components of the electric field. The structure is illuminated by coherent monochromatic waves from both sides.  We further enhance the extraordinary spin by invoking coherent perfect absorption \cite{wan2011,dutta-gupta2012,dutta-gupta12} mediated suppression of scattered light. Coherent perfect absorption (CPA) refers to the nearly perfect absorption of coherent monochromatic light incident on a structure leading to nearly zero reflection and transmission (or scattering) from the structure. This phenomenon is extremely sensitive to system parameters and when at certain parameters we observe this sharp fall in scattering from the system, we refer to it as the CPA dip. Under the condition that the coherent perfect absorption (CPA) dip coincides with one of the coupled mode frequencies, one can ensure that all the incident light energy is coupled to the specified mode \cite{sdgthesis}. Thus the enhancement results from  plasmon-mediated local field enhancement as well as the CPA-mediated total transfer of incident light flux. Both CPA and coupled plasmon resonances are system specific, more so when we want them to coincide. Continuous change in any of the parameters will take the system off one of these resonances thus violating the essential requirement of coincidence for the enhancement of transverse spin. We show that for cleverly chosen parameter set coincidence and hence enhancement of spin is possible.  Note that in contrast to most of the previous studies on transverse spin, we retain the full complex nature of the dielectric function of the metal ensuring losses in propagation of the modes. 

\section{Theoretical Formulation}

\begin{figure}[t]
	\centering
	\includegraphics[width=0.6\linewidth]{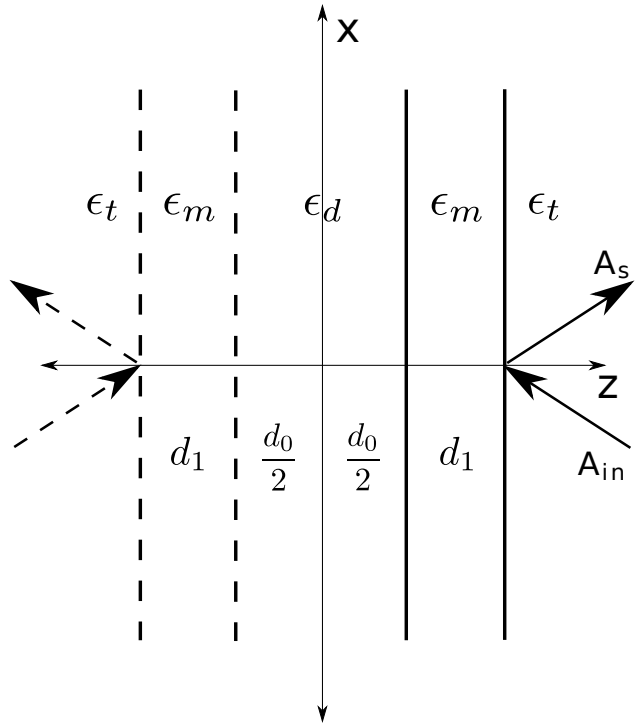}
	\caption{Schematic view of the symmetric gap plasmon guide. We exploit the symmetry and calculate only for the right half of the structure.}
	\label{fig:Geometry}
\end{figure}
Consider the gap plasmon guide shown in Fig. \ref{fig:Geometry}, comprising of metal films of thickness $d_1$ and  dielectric constant $\epsilon_m$  separated by a dielectric gap (spacer layer) of thickness $d_0$ and dielectric constant $\epsilon_d$. The gap plasmon guide is embedded in a dielectric of permittivity $\epsilon_t$. All the media are assumed to be non-magnetic ($\mu = 1$). Let the guide be illuminated from both sides with TM-polarised plane monochromatic light of wavelength $\lambda$ at an angle $\theta$. As mentioned earlier, because of the folded Kretschmann geometry, the light can excite coupled surface plasmons at the inner metal-dielectric interfaces for small enough layer thickness $d_0$. We exploit the symmetry and work only with the right half of the structure. Note that broken spatial symmetry leads to nonreciprocity in reflection \cite{sdg2002,sdg2004,sdg2012} and such a case requires the study of the full five layer structure. 
\par
Following electric-magnetic democracy \cite{berry2009} we write the magnetic field in the dielectric as \cite{sdg2009,sdgbook}:
\begin{equation}\label{eq:0a}
\sqrt{\mu_{0}}H_{y} = \sqrt{\mu_{0}} A_{0}\left(e^{ik_{dz}z}\pm e^{-ik_{dz}z}\right) e^{ik_{x}x}.
\end{equation}
Here and henceforth, the upper and lower signs refer to the symmetric and antisymmetric modes, respectively. Thus, in our case, the symmetry of a given mode is determined by the even/odd character of the magnetic field $H_y$. The corresponding electric field components can be calculated using the Maxwell's equations. For example, for the $x$-component of the field we have:	
\begin{equation}\label{eq:0b}
\sqrt{\epsilon_{0}}E_{x} = p_{dz}\sqrt{\mu_{0}}A_{0}\left(e^{ik_{dz}z}\mp e^{-ik_{dz}z}\right) e^{ik_{x}x} ,
\end{equation}
where $p_{dz}$ is the $z$-component of the scaled wavevector for the dielectric. Similar quantities for the other media are defined as follows:
\begin{equation}\label{eq:0c}
p_{jz} = \frac{k_{jz}}{k_{0}\epsilon_{d}} \ \ \left(j=d,t,m\right)
\end{equation}	
\begin{equation}
k_{jz} = \sqrt{k_{0}^{2}\epsilon_{j} - k_{x}^{2}}	 = i\bar{k}_{jz},~~ k_x=k_0\sqrt{\epsilon_t}\cdot sin\left(\theta\right).
\end{equation}
In Eq.(\ref{eq:0c}) subscripts $d,t,m$ refer to the spacer layer dielectric, ambient dielectric and metal, respectively.

We introduce the abbreviated notations for the scaled electric and magnetic fields
\begin{equation}
\sqrt{\mu_0}A_0 =\tilde{A_0}\ \ \ \  \sqrt{\mu_0}\mathbf{H} =\mathbf{\tilde{H}}\ \ \ \ \sqrt{\epsilon_0}\mathbf{E} =\mathbf{\tilde{E}}.
\end{equation}
Further, we use the boundary conditions (continuity of the tangential components of the electric and magnetic field vectors: $H_y$ and $E_x$ at the two interfaces at $z=d_0/2$ and at  $z=d_0/2+d_1$) to arrive at the matrix equation as follows \cite{sdgbook}
\begin{align}\label{eq:1}
\left(\begin{array}{cc}
1 & \pm 1 \\
p_{dz} & \mp p_{dz} \end{array}\right)
\left(\begin{array}{c}
\tilde{A}_{0} \\
\tilde{A}_{0} \end{array}\right) = 
M_{d_0/2}M_{d_1}\left(\begin{array}{cc}
1 & \pm1\\
p_{tz} & \mp p_{tz}\end{array} \right)
\left(\begin{array}{c}
\tilde{A}_{s} \\
\tilde{A}_{in} \end{array}\right),
\end{align}
where $M_{d_0/2}$ and $M_{d_1}$ are the characteristic matrices \cite{sdgbook} for half of the spacer layer and the metal film, respectively. In Eq.(\ref{eq:1}), $\tilde{
	A}_{in}$ stands for the amplitude of the incident light and $\tilde{A}_s =\tilde{A}_r +\tilde{A}_t$ gives the total scattered amplitude. Here $\tilde{A}_r$ gives the reflected amplitude for light incident from right and $\tilde{A}_t$ corresponds to the transmitted amplitude for left incidence. It is clear from Eq.(\ref{eq:1}) that for a given incident amplitude we can calculate the scattered amplitude as well as the field amplitude $\tilde{A}_0$, and consequently field distributions everywhere in the structure. CPA \cite{wan2011,dutta-gupta2012} refers to the case when we have null scattered amplitude for certain parameters of the system and for specific illumination conditions (angle of incidence, wavelength of light). The suppression of scattering is caused by near-perfect destructive interference of the constituent waves. It corresponds to the case where the amplitude transmission and reflection coefficients match. Thus in the context of the total structure CPA corresponds to the case where $|r_f|=|t_b|$ and there is a phase difference of $\pi$ between $r_f$ and $t_b$ (where $r_f$ is the reflection coeffieicent of the forward propagating wave and $t_b$ is the trasnmission coefficient for the backward propagating wave) \cite{dutta-gupta2012}. Analogous CPA conditions have been exploited for anomalous reflection of light \cite{dutta-gupta12}.
\par
Making use of Eq.(\ref{eq:1}) one can write the explicit expressions for the electric and magnetic fields in the dielectric for the symmetric mode as
\begin{equation}\label{eq:7}
\mathbf{\tilde{H}}_d^s = 2\tilde{A}_0 \cosh\left(\bar{k}_{dz}z\right)e^{ik_x x}\hat{y}
\end{equation}
\begin{equation}\label{eq:8}
\mathbf{\tilde{E}}_d^s = \frac{\tilde{A}_0}{k_0\epsilon_{d}} \left[-2i\bar{k}_{dz} \sinh\left(\bar{k}_{dz}z\right)\hat{x} - 2k_x \cosh\left(\bar{k}_{dz}z\right)\hat{z}\right] e^{ik_x x}.
\end{equation}\\		
The analogous expressions for the antisymmetric mode are:
\begin{equation}\label{eq:9}
\mathbf{\tilde{H}}_d^a = -2\tilde{A}_0 \sinh\left(\bar{k}_{dz}z\right)e^{ik_x x}\hat{y}
\end{equation}		
\begin{equation}\label{eq:10}
\mathbf{\tilde{E}}_d^a = \frac{\tilde{A}_0}{k_0\epsilon_{d}} \left[-2i\bar{k}_{dz} \cosh\left(\bar{k}_{dz}z\right)\hat{x} + 2k_x \sinh\left(\bar{k}_{dz}z\right)\hat{z}\right]e^{ik_x x}.
\end{equation}	
Similar expressions can be obtained for fields in the metallic layer and the ambient dielectric also.
It is then a straightforward job to calculate the spin angular momentum density  $\mathbf{s}$ \cite{berry2009} in all the layers as follows
\begin{equation}
\mathbf{s} = \frac{Im\left(\epsilon_{0}\mathbf{E^*}\times\mathbf{E} + \mu_{0}\mathbf{H^*}\times\mathbf{H}\right)}{4\omega},
\end{equation}
where $\mathbf{H}$ and $\mathbf{E}$ represent the original electric and magnetic fields and not the scaled magnetic fields introduced earlier.
\par
It is important to note that because of the TM character of the incident fields there is no magnetic field contribution to the spin angular momentum density ($\mathbf{H}$ has only one non-zero component along the $y$-direction). The other imporant point to be noted is the phase lag between the $x$ and $z$ component of the electric field vector (see Eqs. (\ref{eq:8}) and (\ref{eq:10})). In both symmetric and antisymmetric modes there is a phase difference between the $x$ and $z$ components of the electric field. As reported by others \cite{bliokh2012,bliokh2014,bliokh2015,bekshaev2015,aiello2015} this gives rise to a polarization ellipse in the $x-z$ plane (plane of incidence). The resultant rotation of the electric field vector leads to the transverse spin. This transverse spin has no reference to the helicity of the light used for excitation and hence this is helicity-independent.
\par

\begin{figure}[t]
	\centering
	\includegraphics[width=0.7\linewidth]{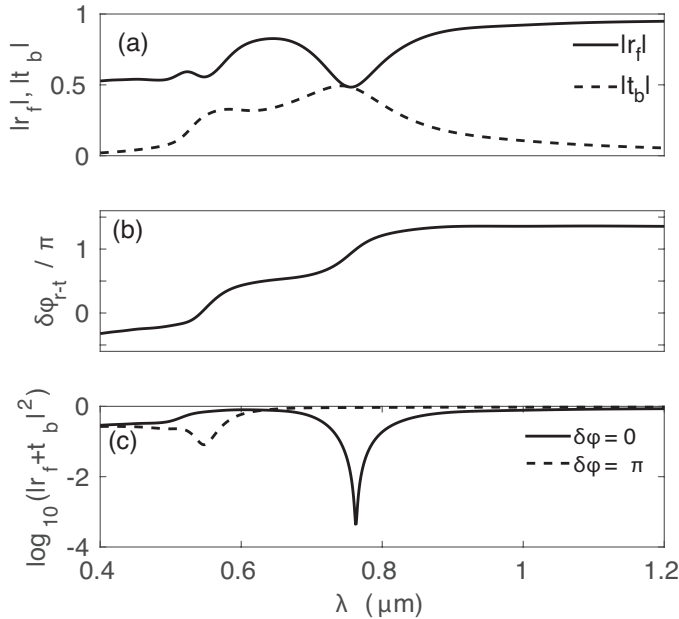}
	\caption{(a) Absolute value of the amplitude reflection (transmission) coefficient $|r_f|$ ($|t_b|$) as function of wavelength $\lambda$, (b) phase difference $\delta \phi_{r-t}$ in units of $\pi$ and (c) the total scattering $\log_{10}|r_f+t_b|^2$ for the gap plasmon guide with $d_0 = 0.473\ \mu m$, $d_1 = 0.05\ \mu m$ and angle of incidence $\theta = \frac{\pi}{4}$. As in \cite{dutta-gupta2012} the subscripts \textit{f} and \textit{b} refer to forward and backward illumination, respectively. The other parameters are as follows $\epsilon_d =1$,  $\epsilon_t = 2.28$ and the value of $\epsilon_m$ is taken from the work of Johnson and Christy \cite{johnson1972}.}			
	\label{fig:Scattering}
\end{figure}

In order to enhance the effect we make use of the selective mode excitation mediated by coherent perfect absorption \cite{sdgthesis}. In fact, the structure under consideration can support both symmetric and antisymmetric modes for sufficiently small $d_0$ and one can tune the parameters of the structure such that CPA is achieved at one of these mode frequencies. It is now understood that there is nearly a $\pi$ phase gap between the split modes \cite{dutta-gupta12}. Thus the CPA realised at one of these coupled modes would correspond to superscattering \cite{sdgbook} at the other coupled mode. In order to have CPA at the other mode, one needs to phase shift one of the incident light waves by $\pi$. Using these features we show numerically that CPA can be achieved at one of these modes ensuring total transfer of incident energy to the specified mode. Thus one can achieve two-pronged local field enhancement - one mediated by the mode of the structure and the other mediated by total absorption of incident light energy. All these features are shown in the next section.
\par

\section{Numerical Results}
In what follows we present the result of our numerical computations. We use the following system parameters: $d_0 = 0.473\ \mu m$, $d_1 = 0.05\ \mu m$, angle of incidence $\theta = \frac{\pi}{4}$, $\epsilon_d =1$,  $\epsilon_t = 2.28$. In some of our calculations we look at the scattering data and the transverse spin as functions of wavelength. To compare data across different wavelengths, we set $\tilde{A}_{in} = \sqrt{\mu_0}A_{in} = 1$ where $A_{in}$ is in units of A$\cdot$m$^{-1}$. 
It must be noted that ultra-thin metal films (with thickness less than about $10$ nm) can have a dielectric constant distinct from the bulk equivalent \cite{maier2015,muralidharan2013,mcpeak2015}. In our calculations we use films of about $50$ nm thickness and bulk-data for the dielectric constant of the metal (gold) from the work of Johnson and Christy \cite{johnson1972} can be used \cite{maier2015, dionne2005}. Note, however, that different data sets do produce different results \cite{dionne2005}.
\par
We first present the results on selective mode excitation mediated by CPA \cite{sdgthesis}. Results are shown in Fig. \ref{fig:Scattering} where we have plotted the results for the total structure. As in \cite{dutta-gupta2012} the subscripts f and b refer to forward and backward illumination, respectively. The results for the absolute values, the phase difference between the reflection and transmission coefficients and the total scattering ($\log_{10}|r_f+t_b|^2$) from the structure  are shown in Figs. \ref{fig:Scattering}(a), \ref{fig:Scattering}(b) and \ref{fig:Scattering}(c), respectively. In order to highlight the efficiency of the CPA process we have used a log scale for the vertical axis in \ref{fig:Scattering}(c). It is clear from the figure that the CPA conditions are met for the symmetric mode at $\lambda = 0.7628 \mu$m, where $|r_f|=|t_b|$ and $\delta\phi_{r-t}=\pi$. As mentioned earlier a phase shift of one of the incident rays by $\pi$ leads to the CPA-like dip at the other coupled mode at $\lambda = 0.5472 \mu$m. This is not a perfect CPA dip due to the fact that the phase difference between the two modes is not exactly $\pi$ \cite{dutta-gupta12}.
\begin{figure}[t]
	\centering
	\includegraphics[width=1.0\linewidth]{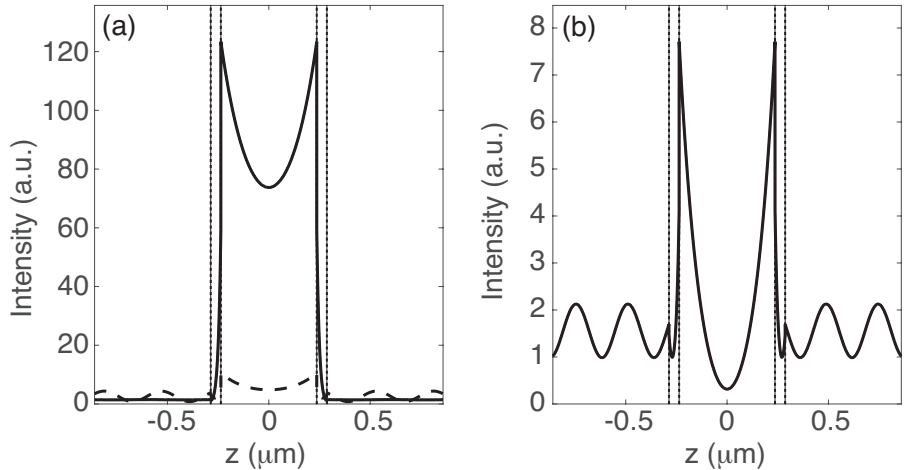}
	\caption{Intensity distribution in the gap plasmon guide for the (a) symmetric and (b) antisymmetric modes at $\lambda = 0.7628~\mu$m and at $\lambda = 0.5472~\mu$m, respectively. The dashed line in (a) gives the profile for $\lambda = 0.7628~\mu$m which is away from the CPA and the mode resonance. Vertical lines depict the boundaries between the various media. The other parameters are as in Fig. \ref{fig:Scattering}.} 
	\label{fig:Intensity}
\end{figure}
\par
\begin{figure}[h]
	\centering
	\includegraphics[width=0.8\linewidth]{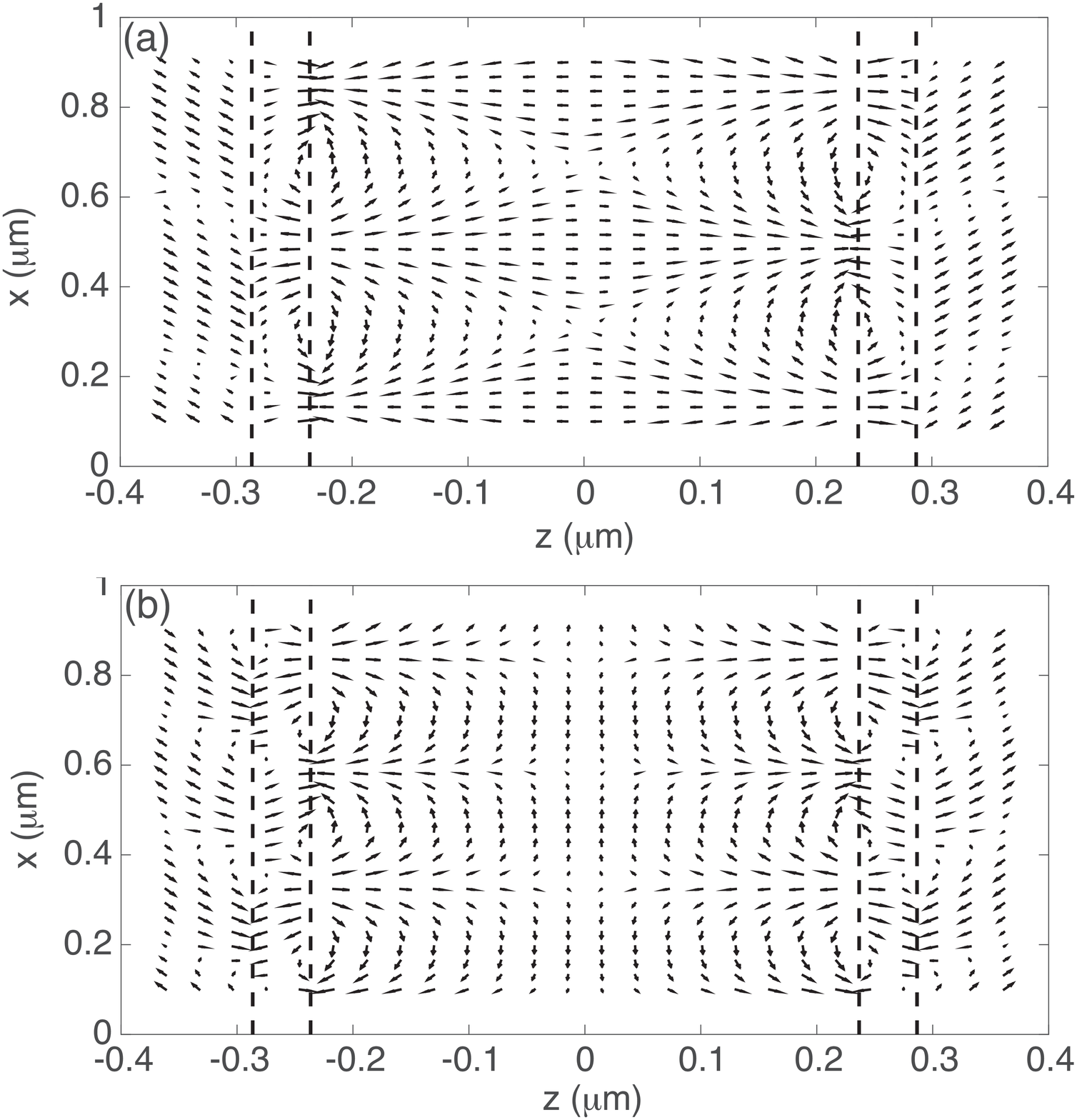}
	\caption{Quiver plot of the electric field for (a) symmetric mode at $\lambda = 0.7628\ \mu$m and (b) antisymmetric mode at $\lambda = 0.5472~ \mu$m. Remaining parameters are the same as in Fig. \ref{fig:Scattering}. Fields are not drawn to scale as the fields in the metal are much smaller than in the dielectric. Dashed vertical lines represent the interfaces between media.}
	\label{fig:Quiver}
\end{figure}
The field enhancement due to CPA is shown in Fig. \ref{fig:Intensity} where the two panels (solid lines) show the intensity distribution in the structure due to the dips at $\lambda = 0.7628 \mu$m and $\lambda = 0.5472 \mu$m, respectively. The dashed vertical lines show the interfaces between the different regions. The total absorption by the structure due to CPA is exemplified by the near null scattered field outside, depicted by the flat background due only to the incident plane. It is interesting to note that a deviation from the CPA condition for the symmetric mode, for example, at $\lambda = 0.7628 \mu$m leads to a drastic drop in the field intensity (see dashed line in Fig. \ref{fig:Intensity}(a)). Deviation from CPA is also associated with the oscillatory character of the field intensity outside due to interference of the incident and the scattered fields. These features are shown prominently for the imperfect CPA dip for the antisymmetric mode at $\lambda = 0.5472 \mu$m (see Fig. \ref{fig:Intensity}(b)). However, mode induced enhancement of the field, though smaller by more than an order of magnitude, can be read from this intensity distribution.

\par	
We now look at the rotation of the electric field vector in the structure as given by Eqs.(\ref{eq:8}) and (\ref{eq:10}) in the dielectric and its analogs in the other media. Results are shown in Figs. \ref{fig:Quiver}(a) and \ref{fig:Quiver}(b), where we have presented the electric field as quiver plots for the symmetric and antisymmetric modes, respectively. Recall that for the symmetric mode we achieve CPA and in the ambient media there are only incident plane waves with near total suppression of scattering and hence the field is not ``structured''. We have only a unidirectional electric field and hence no transverse spin. The same is not true of the antisymmetric mode when there is finite scattering (see Fig. \ref{fig:Intensity}) and one can discern rotations in the field outside due to interference between the incident and scattered light leading to non-zero transverse spin. The scenario is different in the gap where the field is higly structured because of interference of the counterpropagating waves \cite{bekshaev2015}. In this $x-z$ plane, clockwise (counter-clockwise) rotation leads to spin along the positive (negative) $y$-direction. Like in the case of a single interface plasmon \cite{bliokh2012} the transverse spin flips sign across a metal-dielectric interface since the sense of rotation of the field changes.
\par
\begin{figure}[t]
	\centering
	\includegraphics[width=1.0\linewidth]{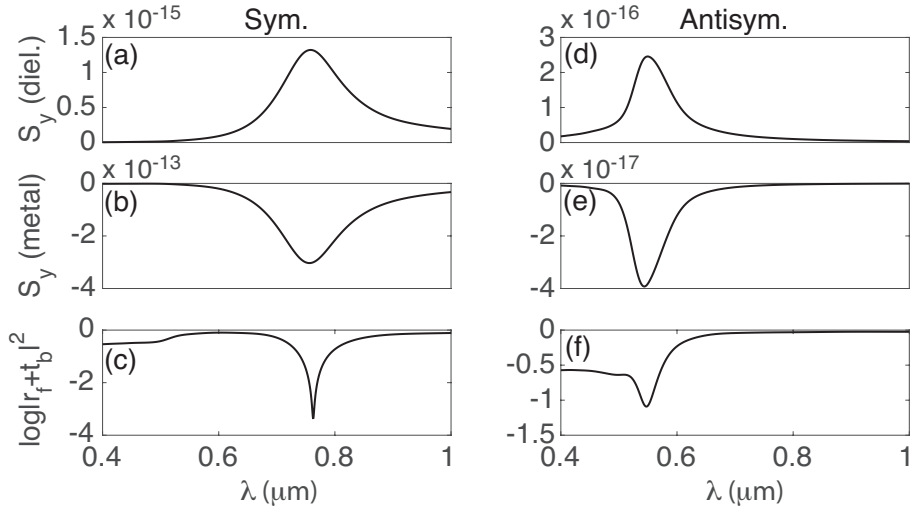}
	\caption{Transverse SAM density as a function of wavelength at the interior metal-dielectric interface from (a) dielectric (b) metal side for the symmetric mode. (d) and (e) give the corresponding quantities for the antisymmetric mode. (c) CPA dip for the symmetric and (f) imperfect CPA dip for the antisymmetric mode. Parameters are as in Fig. \ref{fig:Scattering}. In (a),(b),(d) and (e), the unit of $s_y$ is N$\cdot$s$\cdot$m$^{-2}$ with $A_{in} =1$ A$\cdot$m$^{-1}.$}
	\label{fig:Spin}
\end{figure}
The enhancement of the spin due to the CPA-mediated total transfer of incident energy to the symmetric and the antisymmetric mode is shown in Figs. \ref{fig:Spin}(a), (b) and \ref{fig:Spin}(d), (e), respectively. For reference we have shown the CPA dips in Figs. \ref{fig:Spin}(c) and \ref{fig:Spin}(f). It is clear from Fig. \ref{fig:Spin} that there is an enhancement of the spin with the specified signs in both the cases. As mentioned earlier, the enhancement is due to two factors. The first is the resonant excitation of the modes and its associated local field enhancement. The second is the transfer of incident energy to these modes by means of CPA. Recall that CPA also leads to field enhancement (see Fig. \ref{fig:Intensity}). Since CPA is more effective for the symmetric mode, we have larger transverse spin density for the symmetric mode (compare  the extremal values of $s_y$ in the left and right panels in Fig. \ref{fig:Spin} and note the difference in the vertical scale). It is clear from Fig. \ref{fig:Spin}(a) that at CPA with the symmetric mode at $\lambda=0.7628\mu$m, the transverse spin density in the dielectric is nearly 11.4 times the transverse spin density at $\lambda=0.6\mu$m, which is nearly two full-width at half maximum away from resonance. This clearly indicates that the simultaneous action of the mode resonance and near-total absorption by CPA can lead to a noticeable rise in the transverse spin density that is carried by the plasmon. This point is further demonstrated by the fact that although we have appreciable absorption at lower wavelengths away from CPA (see Figs. \ref{fig:Spin}(c) and (f)), there is no accompanying rise in transverse spin density (see Figs. \ref{fig:Spin}(a), (b) and (d) and (e)) as we do not have CPA mediated mode enhancement at these wavelengths. It is thus clear that the CPA mediated field enhancement adds to enhancing the transverse spin. Moreover, a near-identical width of the resonance features in Figs. \ref{fig:Spin}(a), (b),  and (c) for the symmetric mode is a clear indication that the enhancement is due to CPA combined with the mode excitation.

\begin{figure}[h]
\centering
\includegraphics[width=1.0\linewidth]{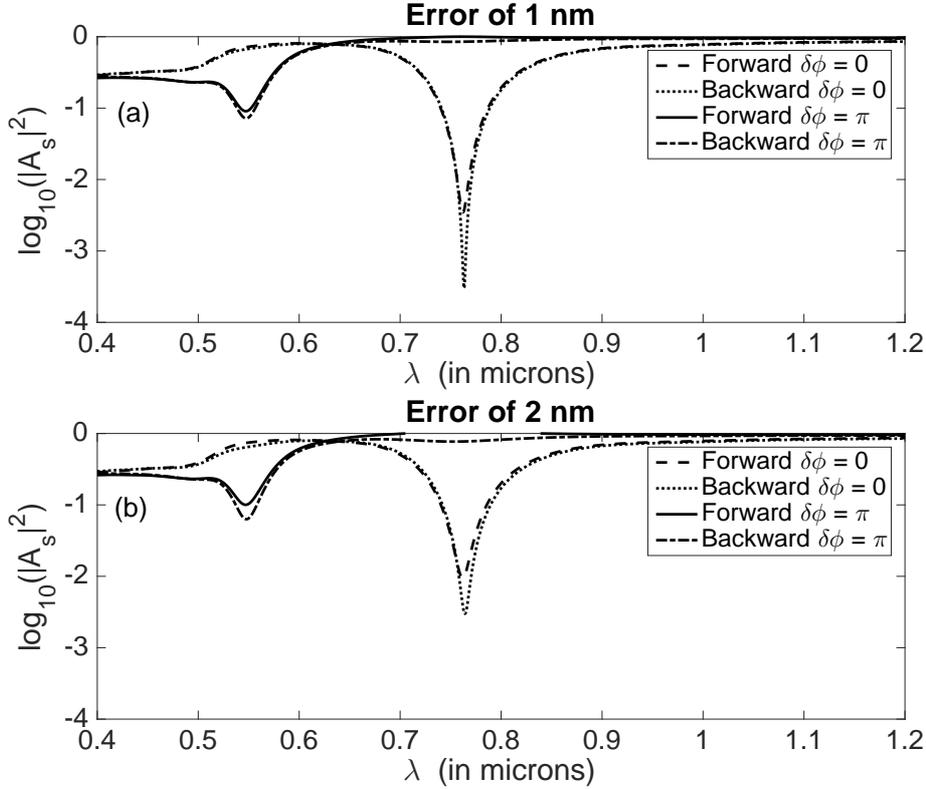}
\caption{Nonreciprocity in light scattered in the left ($|r_f+t_b|^2$) and right ($|r_b+t_f|^2$) half spaces of the asymmetric structure  for (a) $d_{err}$=1 nm and for (b) $d_{err}$=2 nm. The former (latter) is labelled by `forward'(`backward'). Other parameters are the same as in Fig. \ref{fig:Scattering}.}
\label{fig:CPA_Tolerance_Derr}
\end{figure}

\par
\begin{figure}[h]
\centering
\includegraphics[width=1.0\linewidth]{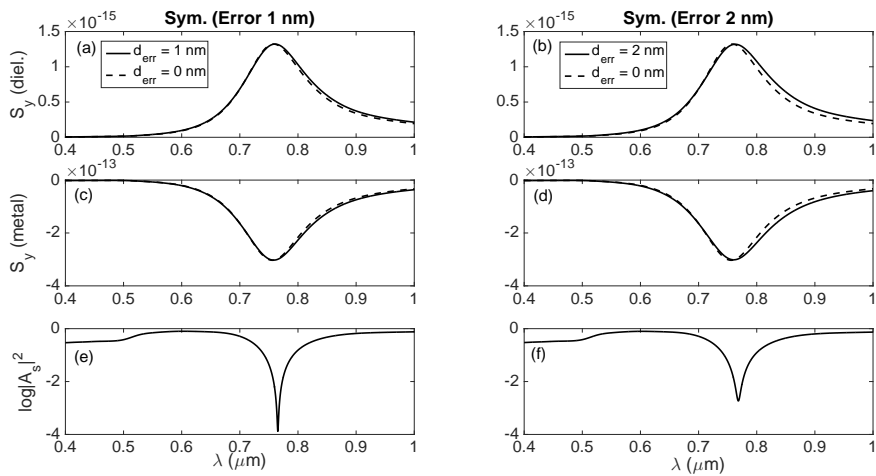}
\caption{Transverse SAM density $S_y$ in (a) dielectric and in  (c) metal, respectively, for  $d_{err}$=1 nm. (b) and (d) are the same as (a) and (c) but now $d_{err}$=2 nm. The dashed lines in (a)-(d) show the transverse spin density when the structure is perfectly symmetric. (e)-(f) show the light scattered from the asymmetric structure on both the sides.}
\label{fig:Spin_Asym_Derr}
\end{figure}

Note that the formulation used in our paper uses symmetry for simplification of the analytical calculations to arrive at closed and compact forms for the fields (see Eqs. (7) -– (10)). It is pertinent to investigate the robustness of the effect against small deviations from the symmetric structure. To this end we study a system where the leftmost (rightmost) layer is made thicker (thinner) by $d_{err}$. It is well known that broken spatial symmetry can lead to nonreciprocity in reflection (and hence scattering) and has been studied in detail \cite{sdg2002,sdg2004,sdg2012}. One then has to distinguish reflection and transmission for left and for right incidence and superpose $r_f$ and $t_b$ on one side, and $r_b$ and $t_f$ for the other side for minimal scattering. It is clear that for a small deviation from the symmetric structure the CPA and the coupled mode resonances will be disturbed. Coupled modes survive in asymmetric structures with different excitation efficiencies for left and right incidence \cite{sdg2012}. However, the broken symmetry may be detrimental for the CPA, since the cancellation of the scattered fields on both sides of the structure could prove to be really difficult. In what follows we show that despite the adverse effect of broken symmetry on CPA, the enhancement in the transverse spin persists. Recall that a proper estimation of the effects of broken symmetry calls for the study of the full 5-layer structure and the analytical expressions for the fields are cumbersome. We invoke the full characteristics matrix approach to calculate the fields and the extraordinary spin. Nonreciprocity features are shown in Fig. \ref{fig:CPA_Tolerance_Derr}, while in Fig. \ref{fig:Spin_Asym_Derr} we have plotted the results for the transverse spin. It is clear from Figs. \ref{fig:CPA_Tolerance_Derr}(a) and (b) that once the symmetry of the structure is disturbed, we have `imperfect' CPA and light can be scattered from the structure. Clearly the amount of light scattered from the structure increases as we increase the asymmetry.
In the context of the transverse spin, it is clear from Figs. \ref{fig:Spin_Asym_Derr} (a)-(d) that even when the symmetry of the structure is disturbed, the enhancement of transverse spin persists. A slight broadening of the feature of the asymmetric structure (as compared to the symmetric case) can be read from the Figs. \ref{fig:Spin_Asym_Derr} (a)-(d) Even though the asymmetry introduced in the system prevents the better realisation of CPA, the incident energy can still be coupled to the plasmon modes leading to enhanced fields and enhanced transverse spin. Thus, the CPA mediated enhancement of transverse spin is a reasonably robust phenomenon.
\par
It is pertinent to note (as stressed earlier) that most of the earlier calculations on transverse spin are carried out ignoring the imaginary part of the dielectric constant. In such a scenario one cannot distinguish between the symmetric and antisymmetric modes because they have no losses though they have different propation constants. In case of coupling through a metal film as in Sarid geometry, this can lead to a distorted picture for the long-range and short-range modes. In fact, the short-range and long-range nomenclature loses its meaning. In contrast, our study retains the full complex character of the dielectric constant.

\section{Conclusion}
\par
In conclusion, we studied a gap plasmon guide under dual plane wave illumination and showed that simultaneous CPA and mode excitation can lead to a noticeable rise in the transverse spin. The enhancement is shown to be robust against small deviations from symmetry. Transverse spin (Belinfante's spin) has been labelled as an elusive quanitity because of the difficulty in its detection. The marriage of CPA with mode excitation in a gap plasmon geometry or in similar resonant structures will facilitate easier observation of this elusive fundamental entity.

\par
\textbf{Acknowledgement.} The authors would like to thank Shourya Dutta-Gupta for help in preparing the manuscript.

\bibliography{References}

\end{document}